\documentstyle[12pt]{article}

\begin{document}

%............................................................................
\def\beq{\begin{eqnarray}}
\def\eeq{\end{eqnarray}}
%............................................................................

%............................................................................
%\mbox{} \hfill ISGBG-03 \\
%............................................................................

\title{\bf Principle of equivalence and wave-paticle duality in 
quantum gravity}

\author
{D. V. Ahluwalia}

\maketitle

\centerline
{Institute for the Studies of the Glass Bead Game (ISGBG)} 
\centerline{Ap. Pos. C-600, Zacatecas 98068, Mexico}
\centerline{Escuela de Fisica, Univ. Aut. de Zacatecas}
\centerline{Av. Preparatoria 301, Fracc. Progreso, Zacatecas 98060, Mexico}

\vskip 2cm
\noindent
{\sc Published in:} {\em Memorias del III Taller de la DGFM-SMF, 2000.
``Aspectos de Gravitaci\'on y F\'isica-Matem\'atica'',} Editadas por
N. Bret\'on, O. Pimentel y J. Socorro.\\

\noindent
{\sc Post-publication Comments:} The factor of $8$ in Eq. (23) should be 
replaced by $4$. This talk is based on: G. Z. Adunas,
E.Rodriguez-Milla, D. V. Ahluwalia, Phys. Lett. B. 485 (2000) 215-223,
and D. V. Ahluwalia, Phys. Lett. A (in press), gr-qc/0002005. The post-talk
evolution of these ideas is contained in those two publications, and in
 G. Z. Adunas,
E.Rodriguez-Milla, D. V. Ahluwalia, Gen. Rel. Grav. (in press), 
gr-qc/0006022.

\newpage
%.............................................................................
\begin{abstract}
{ This talk presents: (a) A quantum-mechanically induced violation of the 
principle of equivalence, and (b) Gravitationally-induced modification
to the wave particle duality. In this context I note that
the agreement between the predictions of general relativity and 
observations of the energy loss due to gravitational waves emitted by 
binary pulsars is just as impressive as the agreement between prediction 
of quantum electrodynamics and the measured value of Lamb shift in atoms.
However, general relativity has not yet yielded to a successful quantised
theory. There is a widespread belief that the two thories are incompatible 
at some deep level. The question is \cite{where}: where? 
Here, I show that the conceptual 
foundations of the  theory of general relativity and quantum mechanics 
are so rich that they suggest concrete modifications into each other in the 
interface region. Specifically, I consider  quantum states that have no 
classical counterpart and show that such states must carry an inherent 
violation of the principle of equivalence. On the other hand, I show that 
when gravitational effects are incorporated into the quantum measurement 
process one must induce a gravitationally induced modification to the de 
Broglie's wave-particle duality. The reported changes into the foundations 
of the two theories are far from in-principle modifications. These are 
endowed with  serious implications for the understanding of the early 
universe and, in certain instances, can be explored in terrestrial 
laboratories.}
\end{abstract}

%..................................... I ..................................
To establish the thesis outlined in the abstract, first
consider a flavor eigenstate of a neutrino
as a standard linear superposition of different mass eigenstates:
\beq
\vert\nu_\ell\rangle=\sum_\jmath {\cal  F}_{\ell\jmath}\,\vert m_\jmath\rangle
\label{nudef}
\eeq
where the flavor index $\ell$  carries the values
$e,\mu,\tau$.  The mass eigenstates are labeled by $\jmath$ which takes on
the  values $1,2,3$. Finally, 
the ${\cal F}_{\ell\jmath}$ are elements of a $3\times 3$ 
unitary mixing matrix, a rough form of which may be
deciphered from the neutrino-oscillation data (see, e.g., ref. \cite{s_dva}).

Equation (\ref{nudef}) makes the non-commutativity of the flavor and 
mass measurements for neutrinos manifest and thus warns for exercising 
conceptual caution. Phenomenologically, the theory of mass and flavor 
eigenstates can be incorporated into the standard model in a generally
covariant manner, i.e.,   in strict accordance with the equivalence principle.
However,  measurement theory does not necessarily respect general covariance 
for  quantum states with no classical analog.  Below, I argue how
the interplay between measurement and the extended standard model 
phenomenology  of equation (1) yields a  subtle violation of 
the equivalence principle.

Now, for simplicity, I confine to a two flavor space.
Let  the ``flavor'' states $\vert \nu_a\rangle$ and $\vert \nu_b\rangle$ 
be the  following linear superposition of mass eigenstates
$\vert m_1\rangle$ and $\vert m_2\rangle$:
\beq
&& \vert \nu_\alpha\rangle = c_\theta \vert m_1\rangle +
 s_\theta \vert m_2\rangle \\ 
&& \vert \nu_\beta \rangle = -s_\theta \vert m_1\rangle +
 c_\theta \vert m_2\rangle
\eeq
where $s_\theta=\sin(\theta)$, and
$c_\theta =\cos(\theta)$.  Without any loss of generality,   
I assume $m_2 > m_1$.
I will first restrict myself to the non-relativistic realm, and then
immediately proceed to the relativistic case. 

%..................................... V ..................................

In  contrast to classical systems, the very quantum construct that 
defines the flavor eigenstates does not allow them to carry a definite mass.  
Therefore, within the orthodox interpretational structure of the theory of 
quantum measurements, the equality of inertial and  gravitational masses 
looses any operational meaning beyond a flavor-dependent fractional
accuracy of the order of: 
\beq
f_\eta:=
\frac{\sqrt {\langle \nu_\eta\vert \hat{m}^2\vert\nu_\eta\rangle
- \langle \nu_\eta\vert \hat{m}\vert\nu_\eta\rangle^2}}
{ \langle\nu_\eta  \vert \hat{m}\vert\nu_\eta\rangle}\label{feta}
\eeq
Here, $\eta=\alpha,\beta$, and  $\hat m\vert m_\jmath\rangle 
= m_j \vert m_\jmath\rangle$, with $\hat m$, the mass operator.
On evaluating $f_\eta$, I find:
\beq
&& f_\alpha
=
 \frac{s_{2\theta}\, \delta m}{2\left(m_1+ s^2_{\theta}\,\delta m\right)}\\
&& f_\beta
=
 \frac{s_{2\theta}\,\delta m}{2\left(m_1+ c^2_{\theta}\,\delta m\right)}
\eeq 
where $\delta m=m_2-m_1$.

%..................................... VI ..................................

Assuming now that both the mass eigenstates carry the same three momentum,
$\vec p$, and within the standard framework of the neutrino oscillation 
phenomenology, I obtain in the relativistic limit: 
\beq
&& f^\prime_\alpha
\simeq\frac{\Delta m^2  c^3 s_{2\theta}}
                     {2\left(2 p^2 c + \Delta m^2  c^3 s^2_{\theta}\right)}
\\
&& f^\prime_\beta
\simeq \frac{\Delta m^2  c^3 s_{2\theta}}
                     {2\left(2 p^2 c + \Delta m^2  c^3 c^2_{\theta}\right)}
\eeq
Here, $\Delta m^2 = m^2_2-m^2_1$ is the parameter that enters
in the kinematic oscillation length:
\beq
\lambda^{osc}_0= \left[\frac{2\pi\,\, \mbox{m}}{1.27}\right]
\left[\frac{E}{\mbox{MeV}}\right]
\left[\frac{\mbox{eV}^2}{\Delta m^2}\right]
\eeq
while $f^\prime_\eta$
is defined as the ratio ${\Delta E_\eta}/{\langle E_\eta\rangle}$ paralleling 
the definition (\ref{feta}). A non-vanishing 
\beq
\Delta f^\prime_{\alpha\beta}&:=&
f^\prime_{\alpha} - f^\prime_{\beta}\simeq
\frac{\left(\Delta m^2\right)^2 c^8}{16 E^4}\,s_{4\theta} \\
&=& 6.25 \times 10^{-26} 
\left[\frac{\left(\Delta m^2\right)^2}{\mbox{eV}^4} \right]
\left[ \frac{\mbox{MeV}^4}{E^4}\right] s_{4\theta}
\label{fprime}
\eeq
introduces an additional length scale into the  neutrino-oscillation 
phenomenology:
\beq 
\lambda^{osc}_{qVEP}=\left[\frac{\pi\,\,\mbox{km}}{5.07}\right]
\left[\frac{10^{-15}}{\vert \Delta f^\prime_{\alpha\beta}\Phi\vert}
\right]
\left[\frac{\mbox{MeV}}{E}\right]\label{lam}
\eeq
In equation (\ref{fprime}) I approximate $E$ by $pc$. This does 
not introduce any error to the indicated order in the mass squared
difference. 
In equation (\ref{lam}), $\Phi$ is an essentially
constant gravitational potential due to the rest of the universe in our 
immediate vicinity. 
To  distinguish better  the classical violation of the equivalence 
principle (VEP), 
from the quantum mechanically induced one,
I here use the abbreviation qVEP for the latter. 
The planetary orbits are not changed by $\Phi$ because it is essentially 
gradient-less over the entire solar system. However, a constant $\Phi$ 
gives rise to significant effects in neutrino physics.
This is because neutrino oscillations form a flavor oscillation
clock, and their evolution is sensitive to
$\Phi$. It is precisely this sensitivity that endows them
with the gravitational red shift dictated by general relativity 
\cite{dva_grf98}
(also see \cite{ac,ac2}). The local cluster of galaxies, referred to as the 
Great attractor (GA), 
embeds the solar system in a few parts in $10^{11}$ constant
contribution to $\Phi$. 
Its value was estimated by 
Kenyon \cite{kenyon}: 
\beq
\Phi_{GA}\simeq 
-3\times 10^{-5}
\eeq 
For comparison, the terrestrial
and solar gravitational potentials on their respective surfaces
are: 
\beq
\Phi_E=-6.95\times 10^{-10},\quad \Phi_S=-2.12\times 10^{-6}
\eeq
The conceptual problems 
associated with estimating $\Phi$ have tempted several authors in the 
neutrino physics community to identify $\Phi_{GA}$ with 
$\Phi$ itself. However, there are no {\em \'a prior\'i} reasons to 
arbitrarily  ignore contributions to $\Phi$ from the sources beyond 
the local cluster of galaxies and to not allow a $\Phi\sim 1$, or even 
larger.

To explore a possible observability of the qVEP for the
long-standing solar neutrino anomaly I combine 
equations (\ref{fprime}) and (\ref{lam}) to obtain:
\beq 
\lambda^{osc}_{qVEP}=
3.16\times 10^{9} \pi \frac{E^3}{\left(\Delta m^2\right)^2 
\vert s_{4\theta} \Phi\vert}\,\, \mbox{km}
\eeq
where it is now implicit that $\Delta m^2 $ is measured in
eV$^2$, and $E$ is expressed in MeV.
>From this one readily obtains the ratio:
\beq 
\frac{\lambda^{osc}_{qVEP}}{\lambda_\odot}
= 0.66\times 10^2 
\frac{E^3}{\left(\Delta m^2\right)^2 
\vert s_{4\theta} \Phi\vert}\label{criterion}
\eeq
where $\lambda_\odot\simeq 1.5\times 10^8\,\,\mbox{km}$ is the mean
Earth-Sun distance relevant for the neutrino detectors on the Earth.

We now study the result (\ref{criterion})
for the solar neutrinos where  $0.2 \le E\le 20\,\,\mbox{MeV}$. 
The existing and planned non-solar neutrino oscillations experiments 
explore $\alpha\le \Delta m^2 \le 10^{-4}$, where $\alpha$ is 
of the order of a few eV$^2$. The neutrino-oscillation 
parameter space, when firmly investigated, can be combined with
result (\ref{criterion}), to check the $E^3$ dependence of qVEP,
and to infer $\Phi$. The
presently controversial  higher end
of the parameter $ \Delta m^2 $ will soon be scrutinized
by an experiment under construction  
at Fermi Lab. in Chicago. The lower end 
of that parameter space is already under intense investigation at
Japan's Super-Kamiokande experiment. The conditions for the observability
of qVEP in the solar neutrino oscillation data requires, (a) that 
the parameters
$\{\Delta m^2,\,\theta,\,\Phi\}$ are  such that the ratio
$\lambda^{osc}_{qVEP}/\lambda_\odot$ is of the order of unity, and 
that (b) the
data quality and analysis be such that they allow for the separation of
the slow and fast degrees of freedom.
The fast degree of freedom comes from the usual kinematically induced 
oscillation length, $\lambda^{osc}_0$,
while the slow one is induced by  qVEP. 
Further, qVEP's $E^3$ dependence can be used to 
distinguish it from the well-known conjecture on VEP due to 
Gasperini \cite{mg,mg2}.

The qVEP is not restricted to neutrino physics.
The Stanford techniques pioneered by Steven Chu \cite{sc}
can be exploited to study qVEP by means of
atomic states  modeled after 
equation (1). At present, these techniques 
allow  for the relevant gravitationally induced phases to be measured with 
the remarkable precision of a few parts in $10^9$.

Having established the phenomena of qVEP, I now present arguments on the
gravitationally modified de Broglie wave particle duality for the
interface region under study.

Some years ago  I formulated an argument
\cite{dva_grf94} that at the Planck scale quantum measurements alter the
local space-time metric in a manner that destroys the commutativity of 
the position measurements. One consequence of that argument was that 
non-locality must be an essential part of any attempt to merge the theory 
of general relativity with quantum mechanics. The non-locality derived in that
argument easily extends to measurements of different components of the position
vector of a single particle, and modifies the fundamental commutators of the
Heisenberg algebra. Efforts in string theories  
come to the same conclusion in an independent manner. In that context 
an early reference is the work of Veneziano \cite{v} while a recent 
one is \cite{sdh}. 
Without invoking extended objects, and entirely within the 
framework of quantum mechanics and the theory of general relativity,
Adler and Santiago \cite{as} have found at the same
modifications to the uncertainty principle as that obtained by works
on extended objects (for directly related works see 
\cite{gac,mm,ns,fs}).
A somewhat different argument, based on the existence of an upper bound
for acceleration, also results in a similar modification to the
uncertainty principle \cite{cls}.
The mathematical expression of the above results that leads to a 
gravitationally
modified expression for the de Broglie wave particle duality is given by
the following modification to the fundamental commutator \cite{ak}:
\beq
\left[{\bf x},\,{\bf p}\right] =i\hbar \left[{\bf 1} +\epsilon \frac{ 
\lambda^2_P {\mathbf p}^2}
{\hbar^2} \right]\label{fun_comm}
\eeq
where $\lambda_{P} = \sqrt{\hbar G/c^3}$, is the Planck length, 
and $\epsilon$
is some dimensionless number of the order of unity. In what follows,
for convenience, I set $\epsilon$ equal to unity. This commutator
reproduces the gravitationally modified uncertainty relations. Specifically,
one has the new position-momentum uncertainty relation:
\beq
\Delta x\,\Delta p_x\ge \frac{\hbar}{2}\left[1+ \left(
\frac{\lambda_{P} \Delta p_x}{\hbar}\right)^2+ \left(
\frac{\lambda_{P} \langle {\mathbf p}\rangle}{\hbar}\right)^2\right]
\label{gmod_ur}
\eeq
which carries with it the Kempf-Mangano-Mann (KMM, ref.\cite{ak})
lower bound on the position uncertainty:
\beq
\Delta x_{_K}=\lambda_{P}\left(1+
\frac{\lambda_{P} \langle {\bf p}\rangle}{\hbar}\right)^{1/2}
\eeq
Notice that $\Delta x_{K}$ has a state dependence via $\langle {\mathbf p} 
\rangle$. For a state of a vanishing $\langle {\mathbf p}\rangle$, one 
obtains 
the absolute minimal distance that can be probed quantum mechanically. 
Since this lowest bound does not depend on the particle species,  
it is likely that this
points towards some new intrinsic property of the space-time itself.

%..................................... XV ..................................

An important implication of the KMM lower bound, $\Delta x_{_K}$, is that
the de Broglie plane waves can no longer represent the physical wave 
functions, even in principle. 
Thus the de Broglie wave particle duality
must undergo a fundamental conceptual and quantitative change.

In their pioneering work KMM have presented a
modification to the de Broglie relation, but they  have 
confined only to the non-relativistic particles (a situation likely to 
be of little interest in the Planck regime). Here I present the 
gravitationally modified de Broglie relation without restrictions 
on the particle's momentum. It is readily seen 
that the momentum space wave function consistent with 
the gravitationally modified uncertainty  relations (\ref{gmod_ur}) 
reads \cite{ak}:

%\vbox{
\beq
\psi(p)&=&
N\left(1+\beta p^2\right)^{-\left[
\,{\kappa({\mathbf p})}/{4\beta(\Delta p)^2}\right]}\nonumber\\
&& \times\exp\left[
-i\,\frac{\langle{ {\mathbf x}\rangle}}{\lambda_{{P}}}
\tan^{-1}\left(\sqrt{\beta} p\right) 
- \frac{\kappa({\mathbf p}) \langle
{\mathbf p}\rangle}{2 (\Delta p)^2\sqrt{\beta}}
\tan^{-1} \left(\sqrt{\beta} p\right)\right]
\eeq
%}
where $\kappa({\mathbf p}):=1+\beta(\Delta p)^2+\beta\langle
{\mathbf p}\rangle^2$, and $\beta:=\lambda^2_P/\hbar^2$. $N$ is a 
normalization factor.
This represents an oscillatory function damped by a momentum-dependent
exponential. I identify the oscillation length with the gravitationally
modified de Broglie wave length:
\beq
\lambda= 2\pi\,\frac{\lambda_{{P}}}{\tan^{-1} \left(\sqrt{\beta} p\right)}
\eeq
Introducing $\overline{\lambda}_{{P}}:={2\pi \lambda_{{P}}}$ as the
{\em Planck circumference}; 
and $\lambda_{{dB}}$ as the gravitationally {\bf un}modified de Broglie
wave length,  $\lambda_{{dB}}=h/p$, the above expression takes
the form:
\beq
\lambda=
\frac{ {\overline{\lambda}_{{P}}} }
{\tan^{-1}\left(\overline{\lambda}_P /\lambda_{{dB}}\right)}
\cases{\rightarrow\lambda_{dB} & for low energy regime  \cr
\rightarrow 4\lambda_{{P}} &
for Planck regime  \cr}
\eeq
In addition, for the specific non-relativistic regime considered 
by Kempf {\em et al.} \cite{ak},
$\lambda$ reproduces their equation (44). This justifies the interpretation
of the oscillatory length associated with KMM's $\psi(p)$ as the
gravitationally modified de Broglie wavelength.

%............................... XVIII .................................

It is worth repeating that in
the Planck realm, the wavelength $\lambda$ asymptotically approaches
the constant value
$4\lambda_{{P}}$ that is now universal 
for all particle species. As a consequence 
of this universality, a new type of coherence is likely to emerge 
in the early universe whose significance for the large-scale uniformity
of the universe has already been mentioned. 

%..................................... XIX..................................

As an illustrative example, 
to the lowest order in $\lambda_{P}$, the effect of the modification 
(\ref{gmod_ur}) on the ground state level of the hydrogen atom results 
in the following modified uncertainty principle estimate for the 
ground state of an electron in an H-atom:
\beq
\left(E_0\right)_g \simeq - \,\frac{m e^4}{2\hbar^2}
\left[
1- \frac{8 m \lambda^2_P}{\hbar^2}\left(\frac{m e^4}{2\hbar^2}\right)
\right]
\eeq
Identifying:
\beq
E_0 = - \,\frac{m e^4}{2\hbar^2}
\eeq
with the ground state level of the hydrogen atom without
incorporating the gravitationally-induced correction to the
uncertainty relation, 
one immediately notices that
the effect of the gravitationally induced modification  
is to reduce the magnitude of the 
ionization energy by $5\times 10^{-48}\,\,\mbox{eV}$. 
This suggests that a space-time endowed with the KMM bound is in some
sense a heat bath as it 
decreases the energy required to disassociate 
the H-atom. 

If the effects of the gravitationally induced modification to the
de Broglie wave particle duality are  
negligible for low energy, their relevance can hardly
be overestimated at the Planck-scale induced phenomena. Apart from the
coherence that is predicted to be present in the early universe, there are 
already speculations that anomalous events around $10^{20}$ eV
cosmic rays may be pointing towards a violation of the Lorentz symmetry 
\cite{lgm,cg}.
This violation is also independently expected from the present study and can
be systematically studied as new cosmic ray experiments yield data at still
higher energies approaching the Planck scale. In this context it is 
important to note that the argument that Planck energy 
cosmic rays are forbidden by the Greisen-Zatsepin-Kuzmin (GZK) cutoff
is no longer valid as the recent cosmic ray experiments have amply 
demonstrated.

%....................... XVII ..............................................

The above discussion makes it clear that the conceptual foundations 
of the  theory of general relativity and quantum mechanics 
are so rich that they suggest concrete modifications 
into each other in the interface region.

%............................ REFERENCES ....................................

%............................................................................

{\bf Acknowledgements.} 
This was my first invited talk in Mexico
after I moved to Mexico as a faculty member
at Zacatecas. I extend my thanks for the invitation, and for the opportunity
to meet a vibrant community full of new ideas and hopes.
My particular thanks go to the leadership of Mexican Physics for making my 
coming to Mexico possible, and so rewarding. 
My editorial 
responsibilities, and efforts involved in settling (not to mention applying
for the CONACyT project and for the SNI membership), proved to be so 
overwhelming for me that so far my Spanish classes have taken a back seat. 
For that my apologies. I hope in some small way I compensate that failing in 
other ways. This work was supported by CONACyT (Mexico) and good cheer of
the desert mountain. Gracias!

\end{document}